\documentclass[twocolumn,eqsecnum,showpacs]{revtex4}
\usepackage{epsfig} 
\usepackage{graphics} 
\usepackage{times}
\usepackage[dvipdfm,colorlinks=true, pdfstartview=FitV, linkcolor=blue, citecolor=blue, urlcolor=blue]{hyperref}
\begin{document}

\title{ Optimization of perturbative similarity renormalization group\\
        for Hamiltonians with asymptotic freedom and bound states}
\author{Stanis{\l}aw D. G{\l}azek}\thanks{ Supported by KBN grant
No. 2 P03B 016 18.}  \author{Jaros{\l}aw M{\l}ynik}
\affiliation{Institute of Theoretical Physics, Warsaw University,
ul. Ho\.za 69, 00-681 Warsaw, Poland}
\date{October 5, 2002}
\begin{abstract}
A model Hamiltonian that exhibits asymptotic freedom and a bound
state, is used to show on example that similarity renormalization
group procedure can be tuned to improve convergence of perturbative
derivation of effective Hamiltonians, through adjustment of the
generator of the similarity transformation. The improvement is
measured by comparing the eigenvalues of perturbatively calculated
renormalized Hamiltonians that couple only a relatively small 
number of effective basis states, with the exact bound state energy 
in the model. The improved perturbative calculus leads to a
few-percent accuracy in a systematic expansion.
\end{abstract}
\pacs{11.10.Gh} \maketitle
\section{Introduction}
The matrix of canonical QCD Hamiltonian is too large for a direct 
diagonalization using computers. One cause of the forbidding size 
is the dynamical coupling among bare degrees of freedom across an 
infinite range of energy scales. This difficulty is manifest in all 
local quantum field theories of physical interest through ultraviolet 
divergences. One possible method to attack the divergence problem 
in the case of asymptotically free theories is to first evaluate an 
effective Hamiltonian and then attempt diagonalization of the 
corresponding effective matrix. For some well adjusted effective 
degrees of freedom may interact over only a limited range of energy 
scales, whereby a definition of a finite effective eigenvalue 
sub-problem becomes conceivable and its subsequent solution using 
computers is not immediately excluded. To derive the effective 
Hamiltonians one can use similarity renormalization group (RG) 
procedure \cite{SGKW1}. This article discusses the accuracy that 
the similarity procedure can achieve in evaluating  the effective 
Hamiltonians in perturbation theory. The latter is the only systematic 
method that exists for handling the initial canonical QCD Hamiltonian.

Having a perturbatively calculated effective Hamiltonian, $\mathcal{H}
(\lambda)$, that strongly couples effective basis states only if their 
free energies, i.e. eigenvalues of certain $\mathcal{H}_0(\lambda)$, do 
not differ by more than a finite width parameter $\lambda$, one can cut 
out a sub-matrix $W_\lambda$ from the matrix of  $\mathcal{H}(\lambda)$ 
in the effective basis. The energy range of the sub-matrix  $W_\lambda$ 
can be limited to few or several widths $\lambda$. Such $W_\lambda$ 
contains only a fraction of  the matrix elements of  $\mathcal{H}(\lambda)$, 
which is why $W_\lambda$ is called a {\it window}. One can diagonalize 
the window numerically and obtain the part of the spectrum of $\mathcal
{H}(\lambda)$ that is matched by the selected window. For example, the 
energy range of the window matrix $W_\lambda$ can be limited to the 
region where a bound state is formed. The question is then how accurate 
the perturbative RG derivation of $\mathcal{H}(\lambda)$ can be for 
the matrix elements in such $W_\lambda$.

The issue was studied previously \cite{SGKW2} using a particularly elegant 
version of the similarity transformation generator, which was taken from
Wegner's flow equation for Hamiltonians in solid state physics \cite{FW}. 
In the model studied in Ref. \cite{SGKW2}, the Hamiltonian $\mathcal{H}
(\lambda)$ could be calculated numerically with arbitrary accuracy. One 
could also calculate the same Hamiltonian in perturbation theory. Then, 
one could compare results of numerical diagonalization of the resulting 
windows cut out from the exact and from the perturbative results for 
$\mathcal{H}(\lambda)$.

The model study \cite{SGKW2} showed that Wegner's generator produced 
quite useful first three terms in an effective coupling constant expansion
for the window Hamiltonian, $W_\lambda = W_0 + g_\lambda \,
W_{\lambda 1} + g^2_\lambda \, W_{\lambda 2}$. Diagonalization of this
$W_\lambda$ produced surprisingly good accuracy on the order of
10\%. The bound state eigenvalue of $W_\lambda$ matched the exact one
with so small error even when $\lambda$ was reduced down to the range
that corresponded to 1 GeV in QCD. Moreover, only a handful of basis states
were sufficient in the numerical diagonalization. However, $g_\lambda$
grew in the asymptotically free model when $\lambda$ was made small, which 
is a generic feature for asymptotic freedom, and $g_\lambda$ became comparable 
with 1 for $\lambda \sim$ 1 GeV. At the same time, the coefficients of 
the perturbative expansion for matrix elements of $W_\lambda$ beyond 
the first three terms turned out to form an alternating series with 
growing coefficients and, eventually, for $\lambda$ still much larger 
than 1 GeV, results including terms proportional to $g_\lambda^3$, 
$g_\lambda^4$, etc. turned out to be useless. Higher-than-second order 
terms led to erratic behavior of $W_\lambda$ and its eigenvalues. This 
result required a closer inspection to decide if one could improve the 
perturbative part of the similarity approach by using other generators 
than Wegner's, and whether any serious investment in a vastly more 
complex QCD calculations should ever be made using this method. The 
problem was not only quantitative, with details depending on the model 
chosen for study, but also qualitative. Namely, one needed to know if 
any significant improvement was possible in principle, because the 
growth of $g_\lambda$ for decreasing $\lambda$ in asymptotically free
theories is in basic conflict with the goal of  reducing the width $\lambda$ 
down to the binding scale using perturbation theory. Since there are 
not many exactly soluble models with asymptotic freedom and bound 
states, it was important to find out if any improvement could be obtained 
in the model of Ref. \cite{SGKW2}.

The optimization task may be attempted in similarity RG procedure 
by taking advantage of the fact that the generator of the similarity 
transformation can be chosen in infinitely many different ways. It is
plausible that the convergence problems emerged in the Wegner case
because interactions with large energy changes were very quickly
eliminated and such rapid elimination could produce large feedback
in the low-energy dynamics, too large for easy reproduction through 
a perturbative expansion. Therefore, this work is focused on the
question if varying the similarity generator, especially the rate at
which the off-diagonal terms are removed, can improve convergence of
at least the first five terms in the expansion, i.e. all terms up to 
and including order $g_\lambda^4$. The number 4 is distinguished by 
the fact that one needs at least four orders of perturbation theory 
to simultaneously account for effective masses and running coupling 
in quark-gluon dynamics in QCD. The answer found in the model studied 
here is positive. Namely, it is shown below that one can obtain a 
systematic expansion for $W_\lambda$ into powers of $g_\lambda$, and 
the sum of first 5 terms in the expansion produces at least three 
times better accuracy than in Ref. \cite{SGKW2}. When an optimized 
version of the similarity generator is identified, one can include 
higher order terms than 4th. For example, it will be shown here that 
6th order improves the accuracy to 10 times better than achieved 
previously using Wegner's generator in the model. Sensitivity of the 
effective theory to the finite bare cutoff in numerical calculations 
is also reduced by at least an order of magnitude (this will be explained 
later) and control over numerical instabilities that emerge in the RG 
evolution is also enhanced (this aspect is only mentioned). The main 
result reported here, however, is not that a class of similarity generators 
leads to improvement in the model chosen for this study, but that the 
flexibility available in the similarity renormalization group procedure 
does lead to considerable options for improved convergence pattern.

Since the similarity approach to QCD has already evolved in the
effective particle version to third order terms \cite{DEG} and
systematic studies of bound states will become possible once the
fourth order terms are brought under numerical control \cite{SGMW}, 
the options for improving convergence need to be further explored 
in more complex models. Also, the development of Wegner's flow 
equations in solid state physics and field theory \cite{AM, AMDC, 
FWPR, SM, SP}, has already demonstrated that new options for 
optimization may have much broader range of applicability than the 
QCD bound state problem could suggest by itself. Besides, if the 
window Hamiltonians of interest could be reliably evaluated using 
perturbation theory, the standard quantum mechanical calculus could 
then be employed to describe with them the time evolution of selected 
states with efficiency probably hard to match by any other method.

This paper is organized as follows. Section \ref{sec:model} describes
the model. Section \ref{sec:pt} discusses results concerning perturbation 
theory and optimization of the similarity generator, which is followed by 
concluding remarks in Section \ref{sec:c}.
\section{Model}
\label{sec:model}
Matrix elements of the model studied here are (see Ref. \cite{SGKW2}
for introduction)
\begin{equation}
\label{Hmn}
H_{mn}=E_{m}\delta _{mn}-g\sqrt{E_{m}E_{n}}\, ,
\end{equation}
where $E_n =b^n$, $b > 1$, and $n$ is an integer, with a convention that 
energy equal 1 corresponds to 1 GeV. The model is cut off in infrared and 
ultraviolet regimes by limiting the subscripts, $M \leq m,n \leq N$, $M$ 
being negative and $N$ positive, both of  much larger magnitude than 1. The 
ultraviolet renormalizability of  the model, its asymptotic freedom, and its 
lack of sensitivity to the infrared cutoff, were discussed in \cite{SGKW2}. 
The bare coupling constant, $g = 0.060600631$, is adjusted to obtain a bound 
state with eigenvalue -1 GeV with 8 digits of accuracy for $b = 2$, $M = -21$ 
and $N = 16$, in exactly the same way as in Ref. \cite{SGKW2}. With these 
choices, the Hamiltonian $H$ is a 38 $\times$ 38 matrix with 37 positive 
eigenvalues, and one negative. The largest energy scale reaches 65 TeV, and 
the smallest one is 0.5 KeV. This energy range saturates the needs of  
contemporary theories where asymptotic freedom and bound states are 
of interest.

The differential similarity RG procedure is based on a class of
equations that can be written as
\begin{equation}
\label{rgw}
\frac{d\mathcal{H}}{d\lambda}=\Big
[F_\lambda\{\mathcal{H}\},\mathcal{H}\Big ] \, ,
\end{equation} 
with the initial condition set at $\lambda = \infty$, forcing the RG
trajectory to start from the initial Hamiltonian, \(\mathcal{H}(\infty)=H \), 
of the bare theory with counterterms. The generator of the similarity
transformation can be written as ($\mathcal{D}_m=\mathcal{H}_{mm}$)
\begin{equation}
\label{czynnik.moj}
\big [ F_\lambda \{\mathcal{H}\} \big ]_{mn} = f_{mn} (\mathcal{D}_m -
\mathcal{D}_n) \mathcal{H}_{mn} \, ,
\end{equation}
where different choices of $f_{mn}$ lead to different results for the
trajectory of effective Hamiltonians. $f_{mn}= \phi_{mn} \, ds/d\lambda$,
where $s=1/\lambda^2$ and $\phi_{mn}\equiv 1$, gives Wegner's
equation, which has a considerable record of applications in solid
state physics \cite{FW, AMDC, AM, FWPR, SM}, independently of the 
similarity renormalization group studies of asymptotically free 
theories. Note also, that multiplication of $\phi_{mn}$ by a 
constant, say $a$, is equivalent to change of variables from 
$\lambda$ to $\lambda'=\lambda/\sqrt{a}$, and corresponds to mere 
shift on logarithmic scale.

\begin{figure}[h]
\begin{flushleft}
\includegraphics[angle=270, width=0.45\textwidth]{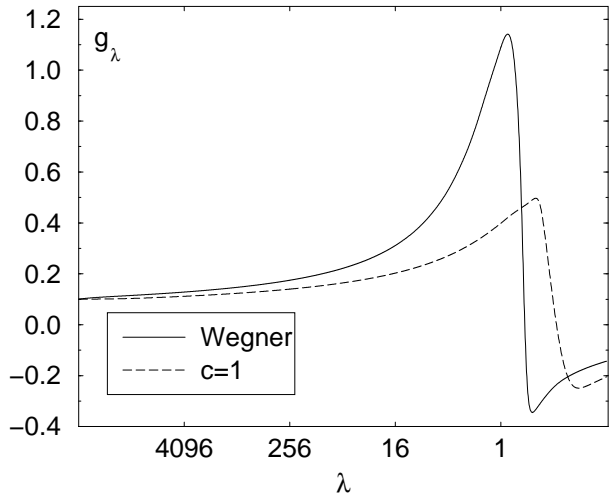}
\end{flushleft}
\caption{The exact running coupling constant $g_\lambda$ from 
Eq. \ref{glambda.definicja} plotted as function of the width $\lambda$. 
The curve labeled ``Wegner'' results from Wegner's equation and the 
curve labeled ``c = 1'' is obtained using generator defined by Eq. 
(\ref{czynnik.moj}) with $\phi_{mn}$ given in Eq. \ref{f_moj}, where 
$c=1$.}
\label{glambdar}
\end{figure}

Behavior of matrix elements $\mathcal{H}_{mn}(\lambda)$ as functions
of $\lambda$ has already been extensively discussed in the case of
Wegner's generator in Ref. \cite{SGKW2}. For $\phi_{mn}$ that is
independent of $\lambda$, it can be {\it qualitatively} represented as
\begin{eqnarray}
\label{qualitative}
\mathcal{H}_{mn}(\lambda ) &  =  &
E_m \delta_{mn}  -  g_\lambda \sqrt{E_m E_n} \nonumber  \\ 
& \times & \exp{\big [-\phi_{mn} (E_m - E_n)^2 /
\lambda^2\big ]} \, ,
\end{eqnarray}
where $g_\lambda$ is the effective coupling constant.  The Gaussian-like 
fall-off function is not exact, but it is not far from the actual form 
factor that makes the interaction Hamiltonian matrix narrow on the energy 
scale, i.e., of width $\lambda$. It is clear that in the case of Eq. 
(\ref{qualitative}) the effective coupling constant could be evaluated 
from the following formula,
\begin{equation}
\label{glambda.definicja}
g_\lambda = 1 - \frac{\mathcal{H}_{M,M}(\lambda)}{E_M} \, .
\end{equation}
The same definition is applied in the case of exact solution. The
resulting effective coupling $g_\lambda$ can be computed numerically
in all cases considered here and its dependence on $\lambda$ is known, 
see Fig. \ref{glambdar}. 
For $\lambda \gg 1$, one has $g_\lambda \sim 1/\log{\lambda}$, and
the growth of $g_\lambda$ toward smaller $\lambda$ continues in case 
of Wegner's generator until the coupling constant slightly exceeds 1, 
which happens when the effective Hamiltonian width becomes comparable 
to the momentum-space width of the bound eigenstate wave function. For
comparison, the optimized generator result is labeled $c=1$. It will 
be discussed later. 

\begin{figure}[h]
\begin{flushleft}
\includegraphics[angle=270, width=0.45\textwidth]{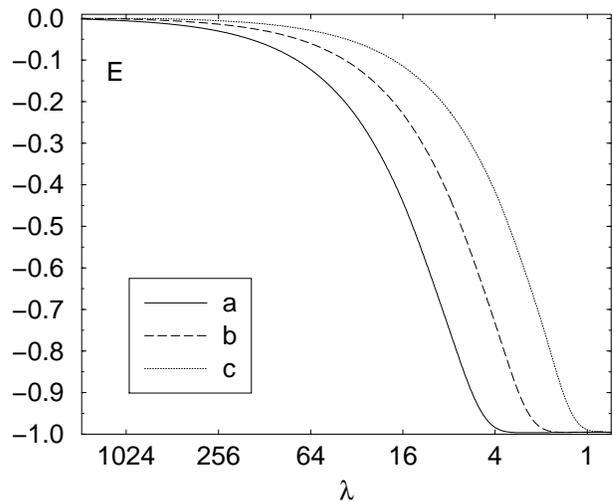}
\end{flushleft}
\caption{Bound state energies obtained by diagonalization of exact
window matrix, $W_\lambda$, with a) $\widetilde{M}=-8 \quad 
\widetilde{N}=2$, b) $\widetilde{M}=-8 \quad \widetilde{N}=1$,
and c) $\widetilde{M}=-8 \quad \widetilde{N}=0$, as functions of
$\lambda$ in case of Wegner's generator.}
\label{wybor_okna}
\end{figure}

Near $\lambda = 1$ GeV the bound state eigenvalue begins to build up on 
the diagonal of $\mathcal H$, with participation of a limited number 
of neighboring matrix elements. The limitation comes from the size of 
$\lambda$ and the actual wave function width. Thus, when one extracts a 
sub-matrix from $\mathcal{H}(\lambda \sim 1)$, which forms a window, 
$W_\lambda$, that embraces all matrix elements that count in the bound 
state dynamics, the diagonalization of that window alone can produce 
the right value of the bound state energy. This is shown in Figure 
\ref{wybor_okna} in case of Wegner's equation for three different 
choices of the window matrix, i.e. for three different choices of  
$\widetilde{M}$ and $\widetilde{N}$ that limit $W_\lambda$ in the same 
fashion as $M$ and $N$ limit $H$ in Eq. (\ref{Hmn}).

Figure \ref{wybor_okna} demonstrates that an exactly calculated small 
window may provide correct results when $\lambda$ is appropriately small, 
i.e. below about 4 GeV. The only problem is that the calculation of a
window Hamiltonian with finite $\lambda $ in realistic theories is not
feasible beyond perturbation theory. Thus, one is led to the question
how well the same small windows can be calculated perturbatively. The
question is of key importance to QCD because the region of $\lambda
\leq 4 $ GeV is where a significant raise of the effective coupling
constant $g_\lambda$ is expected. But the question may also be 
relevant in other areas of physics than strong interactions.
\section{Perturbation Theory}
\label{sec:pt}
This Section describes difficulties that similarity approach encounters 
with convergence of perturbation theory, treating the case with 
similarity transformation generator taken from Wegner's flow equation 
as a benchmark, and shows how one can optimize the convergence by 
changing the generator. The quality of the perturbative expansions
is established in the model by comparing the negative eigenvalue of 
perturbatively calculated windows $W_\lambda$ with the exact value 
of  $-1$ GeV.

Perturbation theory is constructed in the following way. First one
writes a series for $\mathcal{H}(\lambda)$ in the form
\begin{equation}
\label{hlambda.szereg}
\mathcal{H}(\lambda )=\sum_{n=0}^{\infty} g^n \mathcal{H}^{(n)}
(\lambda) \; ,
\end{equation}
where $g$ is the bare coupling constant from $H$ of Eq. (\ref{Hmn}). 
For example, in second order expansion the solution of Eq. (\ref{rgw}) 
with $\phi_{mn}$ independent of $\lambda$, reads
\begin{eqnarray}
\label{H_lambda_2_moj}
\mathcal{H}(\lambda ) 
& = & E_m \delta_{mn}                                             \nonumber\\  
& - & g \sqrt{E_m E_n} 
\exp{\big [-\phi_{mn} (E_m - E_n)^2/\lambda^2 \big ]}       \nonumber\\ 
& + & g^2 \sqrt{E_m E_n} 
\exp{\big [-\phi_{mn} (E_m - E_n)^2/\lambda^2\big ]}        \nonumber\\ 
& \times  & \Bigg[ \sum_{k=M}^{N} E_k \frac{\phi_{mk} (E_m - E_k) - 
                    \phi_{kn} (E_k - E_n)}{A_{mnk}}                  \nonumber\\ 
& \times  & \Big[ 1 - \exp{\big ({-A_{mnk}\over \lambda^2}\big )}\Big]
- 2 \phi_{mn}{(E_m - E_n)^2 \over \lambda^2} \Bigg]        \nonumber\\ 
&  +  & \,\, corrections \, ,
\end{eqnarray}
where
\begin{eqnarray}
A_{mnk} & = & 
\phi_{mk} (E_m - E_k)^2 + \phi_{nk} (E_n - E_k)^2 \nonumber\\ 
& - & \phi_{mn} (E_m - E_n)^2 \, ,
\end{eqnarray} 
and {\it corrections} are of higher order in $g$ than second. The
latter terms are also calculable analytically in orders 3rd, 4th, 
and higher, but they are too complex to display here and hardly 
useful in the analytic form for numerical calculations in
the present study, because they involve combinations of mutually
canceling terms that contain ratios of functions that approach zero in
numerator and denominator and require careful evaluation. Instead of
such analytic expressions, which are discussed here only to
demonstrate what the required expansion consists in, it is more
practical in the model to apply a numerical algorithm and generate
matrices $\mathcal{H}^{(n)} (\lambda)$ order by order through
Runge-Kutta integration of $(N-M+1)(N-M)/2$ coupled non-linear
differential equations.

Once the expansion of Eq. (\ref{hlambda.szereg}) is calculated to some
order, one uses Eq. (\ref{glambda.definicja}) to express $g_\lambda$ 
as a series in powers of $g$ to the same order and one inverts that series 
to obtain $g$ in the form of a new series in powers of $g_\lambda$ up
to that same order. The latter series is then inserted into Eq. 
(\ref{hlambda.szereg}), which provides the desired effective coupling 
constant expansion for the whole matrix $\mathcal{H}(\lambda)$, i.e. 
the expansion in terms of powers of $g_\lambda$. Note that this 
procedure can be carried out without detailed analytic knowledge 
of how the matrices $\mathcal{H}^{(n)} (\lambda)$ depend on $\lambda$,
although such knowledge and selection of relevant, marginal, and irrelevant
matrices may be of great help in calculations with more than one coupling 
constant if one can reliably identify correlations among the couplings. In 
fact, the accuracy achieved here in plain perturbation theory can most 
probably be further enhanced using  techniques similar to the ones
employed in Ref. \cite{SP}. 

\begin{figure}[h]
\begin{flushleft}
\includegraphics[angle=270, width=0.45\textwidth]{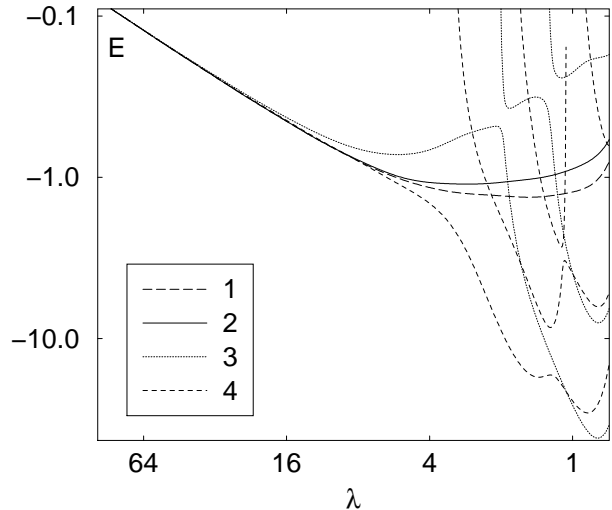}
\end{flushleft}
\caption{Negative eigenvalues obtained from diagonalization of
the window $W_\lambda$ with $\widetilde{M}=-8$ and $\widetilde{N}=2$,
obtained with Wegner's generator in first four orders of perturbation
theory. The orders 3 and 4 produce spurious negative eigenvalues (see
the text for details) with no sign of convergence.}
\label{4wegner}
\end{figure}

The accuracy of  the perturbative procedure for evaluating effective 
Hamiltonians in case of Wegner's equation, i.e. with $\phi_{mn} 
\equiv 1$, is shown in Fig. \ref{4wegner}, in terms of the negative 
eigenvalues of a selected window $W_\lambda$. While the window 
matrix elements are evaluated in perturbation theory, the eigenvalues of 
the window are obtained from non-perturbative diagonalization. Note 
that the vertical axis has logarithmic scale. While the second order  result
produces about 10\% accuracy after diagonalization, the next orders in 
the expansion are wrong by huge factors and completely unacceptable, 
including the fact that they render spurious negative eigenvalues. Since 
it is not known how to select the right negative eigenvalue of the windows 
obtained in third and fourth order calculation, all negative 
eigenvalues of the windows are displayed, labeled with the number that 
indicates the order of the expansion for $W_\lambda$.

The window $W_\lambda$ used in Fig. \ref{4wegner} was selected on the
basis of results analogous to those shown in Fig. \ref{wybor_okna}. An
arbitrarily set requirement was adopted, that a good window should produce
the negative eigenvalue with accuracy around 0.5 percent for $4 \geq
\lambda \geq 1$ when calculated exactly. The 0.5\% may be considered a
large error but it is introduced taking into account that low order perturbation 
theory for $W_\lambda$ is expected to lead to much larger errors. The bound 
state energies calculated by diagonalization of exact windows with different 
values of $\widetilde{N}$ and one value of $\widetilde{M}=-8$ are shown in 
Figure \ref{wybor_okna}. It is clear that one should choose $\widetilde{N} = 
2$ to make sure that the correct result is contained in the window of choice 
when $\lambda \le 4$. For $\widetilde{N}$ kept equal 2 at $\lambda = 2$, one 
obtains the eigenvalue $-9.9599$ with $\widetilde{M}=-8$, $-9.9199$ with
$\widetilde{M}=-7$, and $-9.8398$ with $\widetilde{M}=-6$. Therefore,
$\widetilde{M}=-8$ and $\widetilde{N} = 2$ were selected in Fig. \ref{4wegner}.

The summary of results obtained for $W_\lambda$ from the perturbative
expansion of Wegner's equation is that the expansion produces accuracy
on the order of 10\% in second order but completely fails in higher
orders. Therefore, it is important to verify what happens when one
considers $f_{mn}$ with $\phi_{mn} \neq 1$ in Eq. (\ref{czynnik.moj}). 
Two basic options can be considered. One is to make $\phi_{mn} > 1$, 
which accelerates changes in the matrix elements $\mathcal{H}_{mn}(\lambda)$ 
as functions of  $\lambda$, and the other is to make  $\phi_{mn} < 1$, which 
slows down the changes.

\begin{figure}[h]
\begin{flushleft}
\includegraphics[angle=270, width=0.45\textwidth]{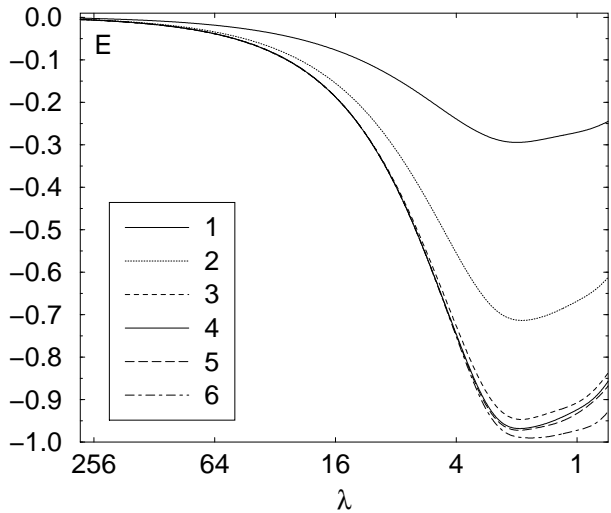}
\end{flushleft}
\caption{Bound state energy obtained by diagonalization  of the same
window Hamiltonian as used in Fig. \ref{4wegner}, but with the
Hamiltonian matrix elements calculated in first six successive orders
of perturbation theory using Eq. (\ref{f_moj}) for the similarity
generator, instead of Wegner's.}
\label{moje}
\end{figure}

Numerical calculations show that the acceleration with a factor of the 
type $\phi_{mn} = 1 +  c \, |m-n|$ with $c$ ranging between 0.01 and 100
does not change the erratic behavior of the window to any better. In
contrast, slowing down the evolution by the factor of the type
\begin{equation}
\label{f_moj}
\phi_{mn}=\frac{1}{1+ c \, \mid m - n \mid} \, ,
\end{equation}
produces significant improvements. The type of results one obtains
from the slowed down similarity transformation is illustrated by an
example in Fig. \ref{moje}, using Eq. (\ref{f_moj}) with $c=1$.
In this case, one obtains only one negative eigenvalue  in the range
$[-2,0]$ (no spurious negative eigenvalues appear in that
range). Moreover, the accuracy of the eigenvalue calculated by
diagonalization of the fourth order window reaches 3\%,  which is
unattainable using Wegner's generator.

The particular choice of $c=1$ follows from the observation that 
4th order results are less accurate than the 3rd order for $c 
\lesssim 0.5$, while for $c \gtrsim 2$ no further improvement 
is obtained in the accuracy of the 4th order terms. Figure \ref{moje}
shows also results of including orders 5th and 6th for $c = 1$.
Inclusion of 6th order contribution achieves the accuracy of  1\% for 
the bound state eigenvalue, which is 10 times better than with 
the Wegner generator. Since the contribution of 6th order is
larger than 5th, one may consider larger $c$, around $2$, to obtain
a more systematic inclusion of successive orders as required. This
issue will not be further discussed here, since already the 4th order 
of rigorous perturbation theory (two loops) for effective Hamiltonians 
in QCD is currently not known.

Finally, one has to address the issue of renormalizability since
for $c \rightarrow \infty$ the similarity transformation generator 
approaches zero within the bounds set by the ultraviolet cutoff $N$
and no RG evolution occurs. The question is then how strongly the 
effective window Hamiltonians $W_\lambda$ for $c \sim 1$ and 
$\lambda \lesssim 4$ GeV, depend on $N$. This can be studied by
inspection of examples and the result is that the generator with
$c = 1$ produces considerably more efficient approach to renormalized
effective theory than Wegner's. The measure of how quickly the 
renormalized effective theory is achieved  is found by evaluating 64 
matrix elements of  the exact window $W_\lambda$ with $\widetilde{M}=-4$ 
and $\widetilde{N}=3$ for $N = 16$, and for $N = 20$. These matrix elements 
are most important in the formation of bound-state in the effective dynamics. 
In both cases the bare coupling constant is adjusted to produce the same bound 
state eigenvalue of -1 (one obtains $g = 0.04878048667$ in the case $N=20$). 
Then, one evaluates 
\begin{equation}
\label{staszek_1}
r^2(\lambda) = { 1 \over (\widetilde{N} - \widetilde{M})^2 } \sum_{m,n = 
\widetilde{M}} ^{\widetilde{N}} \left [ { W_{\lambda m n}(N=16) 
\over  W_{\lambda m n}(N=20)} \,\,-\,\,1 \right ]^2 \, ,
\end{equation}
in the range of $\lambda$ between 4 and 1 GeV where, as indicated by 
Figs. \ref{4wegner} and \ref{moje}, the chosen window is most suitable
for diagonalization. One cannot  rely on a comprison at one value of $\lambda$
because, as it is visible in Fig. \ref{glambdar}, the formation of bound states occurs 
in Wegner's case ($c =0$) and for $c \neq 0$ at different values of $\lambda$, 
being shifted toward smaller values of  $\lambda$ for $c=1$. It turns out that 
the resulting measure $r(\lambda)$ for the Wegner generator varies between 
$10^{-5}$ and $10^{-4}$, with numerical accuracy being of significance, 
while for $c = 1$ one obtains a numerically stable result, approximately equal 
$10^{-6}$ in the whole range between 4 and 1 GeV.

The utility of logarithmic (i.e. depending on exponents $m$ and $n$ of the 
base $b$) slowing down of RG evolution of the off-diagonal matrix elements 
originates from two sources. One is the smaller size of the effective coupling
constant in the region where the bound state is formed, as seen in Fig. 
\ref{glambdar}. For example, at $\lambda = 2$ GeV, in Wegner's case  
$g_\lambda=0.72373722$, and the slowed down case with $c = 1$ gives 
$g_\lambda=0.28518167$, which is about 2.5 times smaller and produces about
40 times smaller result for $g_\lambda^4$. The other source is that
coefficients of the perturbative series in powers of $g_\lambda$ for
matrix elements of $\mathcal{H}(\lambda)$ derived with slower generator are 
not growing with the order as they do in Wegner's case. This is illustrated 
in Tab. \ref{szer_wsp} on two generic examples. Namely, the matrix elements
of coefficient matrices $a_i$ with $i = 0$, 1, 2, 3, and 4, in 
\begin{equation}
\label{szereg4}
\mathcal{H}(\lambda) = a_0 + a_1 g_\lambda + a_2  g_\lambda^2 
                     + a_3  g_\lambda^3 +  a_4 g_\lambda^4 + \,. \, . \, .\, ,
\end{equation}
are tabulated for $(m,n)$ equal $(-1,-1)$ and $(2,1)$ at $\lambda = 2$ GeV,
in Wegner's case and for $c = 1$, as indicated. It is visible that the ratios
$a_4/a_3$ are about an order of magnitude smaller in size for $c=1$ than in 
Wegner's case. 
\begin{table}[!htbp]
\begin{center}
\begin{tabular}{|l|c|c|c|c|c|c|}
\hline 
$\lambda = 2$ & $F\{\mathcal{H}\}$ & $a_0$ & $a_1$ & $a_2$ & $a_3$ & $a_4$ \\
\hline 
$\mathcal{H}_{-1,-1}$ & Wegner & $0.5$ & $-0.498$ & $0.044$ & $0.253$ & $-0.902$ \\
\cline{2-7} 
                                     & $c=1$  & $0.5 $ & $-0.499$ & $-0.689$ & $-1.417$ & $-0.462$ \\
\hline 
$\mathcal{H}_{2,1}$   & Wegner& $0$   & $-1.050$ & $0.270$ & $6.116$ & $-24.788$ \\ 
\cline{2-7} 
                       & $c=1$ & $0$   & $-1.724$ & $-2.685$ & $-2.741$ & $0.960$ \\ 
\hline
\end{tabular}
\caption{Numerical values of perturbative coefficients of successive 
powers of $g_\lambda$ in Eq. (\ref{szereg4}) at $\lambda = 2$ GeV for 
selected matrix elements in case of Wegner's generator and for $c = 1$. 
See the text for details.}
\label{szer_wsp}
\end{center}
\end{table}
\section{Conclusion}
\label{sec:c}
Similarity renormalization group approach to solving asymptotically
free theories could employ various generators of the similarity
transformation to calculate effective Hamiltonians. The calculation
can be carried out in perturbation theory and then the Hamiltonians
can be diagonalized  using computers with a reasonable chance for
obtaining accurate answers, if the perturbative procedure for
evaluating effective Hamiltonians is sufficiently precise. It was
shown here, using Wegner's generator as a benchmark, that in order to
obtain stable 3\% accuracy in the bound state spectrum one needs to
slow down the rate of elimination of the off-diagonal matrix elements.
In the model study, the rate of change of bare Hamiltonian matrix 
elements with largest energy changes was slowed down 38 times.
The stabilizing effect of slowing down the RG flow is evident 
from comparison of Figs. \ref{4wegner} and \ref{moje}. Perturbation 
theory including sixth order terms with $c=1$ in Eq. (\ref{f_moj}) 
achieved  1\% accuracy  in evaluating effective dynamics, as measured 
by the bound state eigenvalue.

However, there is a stiff price to pay. Namely, the second order
perturbative expansions with slower generators produce only about 30\%
accuracy, in comparison to 10\% in Wegner's case, and to take
advantage of the improved convergence one has to go to higher orders,
which requires considerably more effort. On the other hand, orders
higher than second are not attainable using Wegner's generator, and
the outstanding accuracy of about 10\%, achievable already in the
second order, cannot be improved by including higher orders in a plain
expansion in one running coupling constant. Moreover, the slower 
generator with $c=1$ leads also to less sensitivity of the numerically 
evaluated effective theory to the bare cutoff, when the latter is kept 
large but finite.

Detailed studies will disclose how much work is actually required to
achieve the few-percent accuracy with realistic Hamiltonians. However,
the model study described here shows already that optimization of the
similarity generator may in principle lead to a desired convergence
pattern in the case of bound states in asymptotically free theories. 
Perhaps, the optimization found here may turn out to be useful also in 
other applications of similarity than QCD.

\end{document}